\documentclass[epjCONF, onecolumn]{svjour}
\usepackage{graphics,natbib}
\usepackage[varg]{txfonts} 
\usepackage[latin1]{inputenc}
\session-title{Assembling the Puzzle of the Milky Way}
\def\hi{H{\sc i}}
\def\Gyr{\,{\rm Gyr}}
\def\yr{\,{\rm yr}}
\def\cm{\,{\rm cm}}
\def\Tvir{T_{\rm vir}}
\def\kpc{\,{\rm kpc}}
\def\kms{\,{\rm km\,s}^{-1}}
\def\msun{\,{\rm M}_\odot}
\def\lsim{\lower.7ex\hbox{$\;\stackrel{\textstyle<}{\sim}\;$}}
\begin{document}
\title{Accretion by the Galaxy}
\author{James Binney\inst{1}\fnmsep\thanks{\email{binney@thphys.ox.ac.uk}} \and 
Filippo Fraternali\inst{2,3}}

\institute{University of Oxford, Rudolf-Peierls Centre for Theoretical Physics,
Keble Road, Oxford OX1 3NP, UK 
\and University of Bologna, Department of Astronomy, via Ranzani 1, 40126, Bologna, Italy
\and
Kapteyn Astronomical Institute, Postbus 800, 9700 AV, Groningen, NL}
\abstract{
Cosmology requires at least half of the baryons in the Universe to be in the
intergalactic medium, much of which is believed to form hot coronae around
galaxies. Star-forming galaxies must be accreting from their coronae. \hi\
observations of external galaxies show that they have \hi\ halos
associated with star formation. These halos are naturally modelled as
ensembles of clouds driven up by supernova bubbles. These models can fit the
data successfully only if clouds exchange mass and momentum with the corona.
As a cloud orbits, it is ablated 
and forms a turbulent wake where cold high-metallicity gas mixes with
hot coronal gas causing the prompt cooling of the latter.
As a consequence the total mass of \hi\ increases.
This model has recently been used to model the Leiden-Argentina-Bonn survey of
Galactic \hi. The values of the model's parameters that are required to model
NGC\,891, NGC\,2403 and our Galaxy show a remarkable degree of consistency,
despite the very different natures of the two external galaxies and the
dramatic difference in the nature of the data for our Galaxy and the external
galaxies. The parameter values are also consistent with hydrodynamical
simulations of the ablation of individual clouds. The model predicts that a
galaxy that loses its cool-gas disc for instance through a major merger
cannot reform it from its corona; it can
return to steady star formation only if it can capture a large body of cool
gas, for example by accreting a gas-rich dwarf. Thus the model explains how
major mergers can make galaxies ``red and dead.''
} 
\maketitle
\section{Introduction}
\label{intro}
{\bf The Milky Way} is typical of the galaxies that now dominate the cosmic
star-formation rate (SFR): its luminosity lies extremely close to the
characteristic luminosity $L^*$ of the Schechter galaxy luminosity function,
so it is one of the most massive galaxies that are still actively forming
stars. The colour-magnitude diagram of the local stars implies that the
SFR in the thin disc, which is the Galaxy's dominant
component, has declined only by a factor of a few over the last $10\Gyr$
\citep{Gilmore,AumerB}. Since the SFR must depend strongly on the cold-gas
content of the disc, this finding indicates that despite turning
$\sim5\times10^{10}\msun$ of gas into stars, the disc has only mildly
depleted its cold-gas content, which is significantly less than $10^{10}\msun$. It follows
that the Galaxy must be somehow replenishing its supply of cold gas.

Here we present a model of how the Galaxy acquires cold gas. This model was
developed to explain observations of the \hi\ content of nearby galaxies and
has been recently applied to the qualitatively different data for our Galaxy.
Although  the model is still rather crude, and
a number of aspects
need to be worked out in greater detail, it ties together
disparate data in a remarkably coherent manner, which suggests that the
underlying physical picture is correct. 


\section{Cosmological context}\label{sec:cosmos}

\subsection{Missing baryons}
 
Within the context of standard cosmological theory we can estimate the cosmic
density of baryons in two distinct ways \citep[e.g.][]{Peacock}: (i) by combining cosmic
nucleosynthesis with measurements of the abundances of helium, deuterium and
lithium, and (ii) from the power spectrum of
fluctuations in the temperature of the cosmic microwave background. Method
(i) yields the number of CMB photons per baryon, while method (ii) yields the
fractions of the closure density contributed by baryons and by matter as a
whole. Both methods point to baryons contributing $\sim4\%$ of closure
density and method (ii) implies that baryons contribute $17\%$ of the matter
density. 

The gravitating mass of the Local Group can be reliably estimated from either
the classical Kahn-Woltjer timing argument \citep[e.g.][]{BT08} or from a new
formulation of it \citep{Whitexx}, so we know that $\sim8 \times 10^{11} \msun$ of baryons were
originally associated with the Local Group. Observations of the stars and gas
in the Local-Group galaxies give a total baryon content of
$<1.5\times10^{11}\msun$, less than a fifth of the expected amount.

It is generally believed that the missing baryons are contained in hot,
diffuse gas. Unless this gas has escaped from the gravitational field of the
Galaxy and the Local Group, its temperature must lie in a narrow range around
the virial temperature $\Tvir$ of the confining gravitational well --
at temperatures slightly smaller than $\Tvir$, the gas is so narrowly confined
to the centre of the well that its pressure near the Sun would exceed the
pressure within clouds of well-studied cool gas. Gas with
$T\simeq \Tvir$ will be referred to as ``coronal gas''.

In a rich cluster of galaxies the gravitational field is strong enough to
compress coronal gas to densities at which it can be readily detected either
through its X-ray emission \citep{Helfandxx} or through distortion of the
spectrum of the CMB via the SZ effect \citep{SZ72}. In massive
galaxy groups coronal gas can be detected through its X-ray emission, but in
low-mass groups such as the Local Group the density and temperature of
coronal gas is expected to be too low for detection in either X-rays or the
SZ effect to be possible. However, we do have indirect indications of the
existence of coronal gas within the Local Group. One indication is provided
by ultraviolet observations of objects at high Galactic latitude, which
reveal absorption by highly ionised species such as O{\sc vi} along lines of
sight that pass close to \hi\ clouds \citep{Sembach}; models of the interface
between coronal gas and \hi\ predict the existence of regions rich in O{\sc
vi}. In fact, long before the emergence of modern cosmology \cite{Spitzer}
inferred the existence of coronal gas from the existence of absorption lines
of interstellar Na and Ca in the spectra of high-latitude stars by arguing
that without a confining medium at $\sim\Tvir$ the clouds responsible for
this absorption would quickly dissipate. 

Another pointer to coronal gas is the gross asymmetry between the trailing
and leading Magellanic Streams -- the Stream is almost certainly formed of
gas stripped from the SMC-LMC system. If clouds within the Stream were
orbiting freely, the leading stream would be as extensive as the trailing
stream. In fact it is much shorter than the trailing stream \citep{Putman}, a
phenomenon that arises naturally if clouds of \hi\ experience hydrodynamical
drag as they move through ambient gas of density $\sim 10^{-4}\cm^{-3}$
\citep{BenMoore,Harvard}. The morphology of compact high-velocity \hi\ clouds
provides a similar argument for the existence of a low-density ambient
medium: when well resolved, these clouds generally have a tadpole-like
head-tail morphology indicative of distortion by motion through an ambient
medium \citep{Braun}. Thus many lines of argument suggest that the missing
baryons are contained in gas at $\sim2\times10^6\,$K that fills much of the
Local Group.

\subsection{Evidence for accretion}

The history of a galaxy's star formation is encoded in the colour-magnitude
diagram of its stars, which can be probed either directly when the galaxy is
close enough to resolve its stars, or indirectly through the Galaxy's
colours. The conclusion of many studies of CMDs and galaxy colours is that
in spiral galaxies like the Milky Way, the SFR is only slowly declining.
Only $\sim10\%$ of the masses of the discs of  such galaxies is contributed
by detected gas, so much more gas has been turned into stars than the discs
currently possess. Hence either star-formation is about to cease in these
systems, or they are constantly replenishing their gas supply. Since the SFR
of a disc must depend on how much cold gas it has, the slowness of the
decline in the SFR argues against a rapid drop in the stock of cold gas, and
thus in favour of continuous replenishment. 
This conclusion is strongly
reinforced by models of the chemical enrichment of our own disc, which
require constant accretion of relatively metal-poor gas to explain why there
are not more metal-poor stars near the Sun than are observed \citep[the
``G-dwarf problem''][]{PagelPatchet}.

\subsection{Why isn't the Galaxy red and dead}

To understand the impact of the hot IGM on disc galaxies it is natural to
turn to rich clusters of galaxies, where the IGM can be directly observed.
The cooling time of the IGM, being inversely proportional to the density, is
shortest at the centre of the system, so this is where we expect to find the
coolest gas. X-ray observations bear out this expectation, but with a
surprising twist: although near the centre the cooling time is often much
shorter than the Hubble time, there is very little gas at temperatures lower
than $\sim\Tvir/3$ \citep{Peterson03}. Indeed, there is an almost complete
absence of spectral lines from species such as Fe{\sc xvii} that should form
as gas cools through $\sim10^6\,$K. Thus near the centres of rich clusters,
gas in radiating, but not cooling. It follows that its radiative losses are
being offset by a heat source, and the obvious mechanism is mechanical
heating by the bipolar outflows that are known to accompany accretion onto
compact objects, in this case the black holes that reside at the
centres of galaxies.  In fact, mechanical heating by black holes was
predicted by \cite{BinneyT} before observations revealed the absence of cool
gas. Unlike alternative heating mechanisms \citep{Ciottixx}, it is naturally
self-regulating \citep{OmmaB,Binney05}. Regardless of what stops the
temperature of gas in clusters falling below $\Tvir/3$, this phenomenon poses
a puzzle in the context of disc galaxies because it suggests that the black
holes at the centres of spiral galaxies should be able to prevent coronal gas
cooling effectively, just as the black holes at the centres of clusters do.
Indeed, the weakness of the X-ray emission around spiral galaxies compared to
that from rich clusters bears witness to the relative ineffectiveness of
radiative cooling in the coronal gas of spiral galaxies compared to that in
rich clusters. Hence if mechanical heating is effective in rich clusters, why
would it fail around disc galaxies? 

A suggestion that the black hole at the centre of the Galaxy {\it does}
effectively heat the coronal gas is provided by the fact that the bulk of the
star formation in the Galaxy occurs not near the Galactic centre, where the
pressure and density of the coronal gas must be largest and therefore its
cooling time must be shortest. Instead the bulk of star formation occurs
several kiloparsecs away from the centre, in the disc. Moreover, what star
formation does take place near the Galactic centre, in the nuclear molecular
disc, is likely fed by gas that has been driven in by the Galactic bar, from
its corotation resonance to its inner Lindblad resonance. Thus 
we need to
explain why galaxies at the centres of rich clusters long ago ran out of cold
gas despite being enveloped in enormous quantities of strongly radiating
coronal gas, while spiral galaxies manage to stay youthful and blue by
replenishing their stocks of cold gas from their diffuse coronal gas. In
particular we have to explain why in spiral galaxies cooling coronal gas
falls onto the disc several kiloparsecs away from where the cooling time of
the coronal gas is shortest.

\section{Extraplanar HI}\label{sec:extrap}

Sensitive studies of nearby spiral galaxies in the 21-cm line of \hi\ reveal
that significant quantities of \hi\ lie more than a kiloparsec above or below
the equatorial planes of many galaxies \citep[][and references
therein]{Sancisi08}. In edge-on galaxies such as NGC\,891 this conclusion 
follows rather directly from the projected distribution of \hi\ emission on
the sky (although kinematic information is required to rule out the
possibility that the material seen above and below the plane lies in an
extended warped disc). In galaxies that are not seen edge-on, the existence of
gas above and below the plane must be established by detailed kinematic
modelling \citep{Boomsma}. \hi\ that lies more than a kiloparsec from the
equatorial plane is called ``extraplanar \hi'' and constitutes an ``\hi\
halo''.

Studies of extraplanar \hi\ reveal three indications that extraplanar \hi\ is
associated with star formation: (i) the radial extent of \hi\ is similar to
that of significant star formation; (ii) the mass of the extraplanar \hi\ is
correlated with the galaxy's total SFR; (iii) the galaxy's main \hi\ layer
reveals small holes, and these holes are often associated with not only
the signs of recent star formation, but also nearby \hi\ emission at
anomalous velocities. The conclusion is inescapable that localised bursts of
star formation in the disc give rise to supernova bubbles that blast nearby
\hi\ out of the disc and up into the \hi\ halo.

From kinematic modelling of \hi\ halos one can show that the azimuthal velocity
of the \hi\ drops quite steeply with distance from the plane
\citep{Oosterloo07}. This fact proves to be of considerable importance.

\section{A model of extragalactic fountains}\label{sec:fountain}

The idea that star formation will drive a circulation of gas in galaxies goes
back a long way and has been explored by many authors
\citep{Bregman,Collins}. \citealt[(\hskip-3pt][hereafter FB06)]{FraternaliB06} modelled the
\hi\ datacubes of NGC\,891 and NGC\,2403 as emission from an ensemble of
clouds moving on ballistic trajectories. Their clouds are launched from
platforms in the disc that are on circular orbits with a radial distribution
that reflects the distribution of star formation. The clouds' velocities are
of order $h_v$ and have an angular distribution that has to be sharply
concentrated around the Galactic poles. They found that with $h_v\sim75\kms$ this
model was able to account for most aspects of the data for both galaxies,
despite the galaxies' strongly contrasting masses and inclinations. However,
the fit to the data was deficient in two respects: the azimuthal streaming
velocity of the \hi\ halo was predicted to decline too slowly with distance
from plane, and the data indicated a stronger bias towards inflow than the
models could reproduce, even when clouds were assumed to be ionised and
therefore invisible as they were shot upwards. \citealt[(\hskip-3pt][hereafter
FB08)]{FraternaliB08} showed that these deficiencies were removed if the
orbits and masses of clouds were modified by interaction with coronal gas.
The successful models assumed that clouds sweep up the coronal gas they
encounter on their paths, the rate of accretion being quantified by the
e-folding time $\alpha^{-1}$ of a cloud's mass. Because the corona was assumed to be non-rotating,
the accretion of this gas slows the azimuthal motion of clouds, as the data
require. The accretion has two other beneficial effects: it causes infall to
predominate over outflow as the data require, and it prevents net transfer of
angular momentum to the corona by ram pressure. The second point is important
because the moment of inertia of the part of the corona near the optical
galaxy is small, and if not counteracted by accretion, ram pressure would
quickly set this part of the corona rotating nearly as fast as the disc. Then
the action of the disc on the corona would be that of a centrifugal pump. The
values $h_v=75\kms$, $\alpha=1.5\Gyr^{-1}$ provided good fits to the data for
both galaxies. The models predicted net accretion rates of 2.9 and
$0.8\msun\yr^{-1}$ for NGC\,891 and NGC\,2403, respectively, quite similar to
the galaxies' star-formation rates.

\cite{Marinacci10} illuminated the results of FB08 by simulating the motion
of an \hi\ cloud through coronal gas. They found that the cloud was
inevitably ablated by the hot gas flowing over its surface. However, if the
density of the coronal gas and the metallicity of the cloud were high enough,
the total mass of cool gas, visible as \hi, would nonetheless increase
because in the cloud's turbulent wake ablated cool gas would cause coronal
gas to cool and become neutral. Consequently, they argued that the model of
FB08 was fundamentally correct providing their ``clouds'' were understood to
be clouds plus wakes of cool gas. They pointed out that in external galaxies
\hi\ observations lack the sensitivity to detect an individual stream of the
type they predicted, but the \hi\ datacube of our Galaxy does contain
suitably elongated features. \cite{Marinacci11} presented more elaborate
simulations of clouds moving through coronal gas and showed that there {\it
is} a net transfer of momentum from the cloud to the corona if their relative
velocity exceeds a threshold $\sim50-85\kms$ that increases with the coronal
density. The existence of this threshold velocity suggests that coronae spin
at a non-negligible rate, but slower than the disc by $80-120\kms$.

\section{Modelling the LAB datacube}\label{sec:Galaxy}

\cite{Marasco11} applied the model of FB08, modified to include the insights
gained by Marinacci et al.\ (2010, 2011), to the datacube from the
Leiden-Argentine-Bonn (LAB) all-sky survey of Galactic \hi\
\citep{Kalberla05}.  From these data, which are dramatically different in
type from those from which the model was developed, they re-determined the
model's parameters, $h_v$ and $\alpha$. They found $h_v=70\kms$ in line with
the values required by external galaxies, and $\alpha=6.3\Gyr^{-1}$, a factor
$\sim4$ larger than that favoured by external galaxies. Most of this
difference is attributable to the adoption by FB08 of a non-rotating corona:
when the Galaxy's corona is assumed to be non-rotating, the LAB data require
$\alpha=2.5\Gyr^{-1}$.  However, for a rigorous comparison of results for our
galaxy and external galaxies the current model needs to be fitted to the data
for external galaxies.

The key differences between the model fitted to the LAB data and the model
used by FB08 are (i) the adoption of a rotating corona, and (ii) a more
sophisticated parametrisation of how gas is  ionised as it leaves the disc
and neutral on its return: gas becomes neutral when its vertical velocity is
smaller than its value on ejection by a factor $1-f_{\rm ion}$. The value of
$f_{\rm ion}$ is not tightly constrained by the data, but values $f_{\rm
ion}\simeq0.3$ work well, so gas becomes neutral 
quite soon after it is ejected.

From the model that fits the LAB data we can read off the global properties
of the Galaxy's \hi\ layer. The rotation speed falls $\sim20\kms$ below the
circular speed $1\kpc$ above the plane and then by a further $\sim10\kms$
with each kiloparsec. A similar profile was obtained for NGC\,891 by
\cite{Oosterloo07}. The scaleheight of the \hi\ layer increases strongly with
Galactocentric distance $R$ from $\sim0.35\kpc$ at $R=4\kpc$ to $1.75\kpc$ at
$R=12\kpc$. The rate of accretion of pristine gas rises from near zero at
$R<3\kpc$ to a peak value $\sim7\times10^{-3}\msun\yr^{-1}\kpc^{-2}$ at
$R=9\kpc$ and then falls to zero at $R>13\kpc$. Overall the Galaxy is
predicted to accrete $1.6\msun\yr^{-1}$ of \hi\ or $2.3\msun\yr^{-1}$ of gas
when helium is included. For comparison, the Galaxy's star-formation rate is
$\sim3\msun\yr^{-1}$ \citep{Diehl06}, of which $\sim1\msun\yr^{-1}$ will be
provided by dying stars. In addition to gas accreted through the modelled
fountain clouds, \cite{Sancisi08} estimate that accretion of high-velocity
clouds, which are not accounted for by the model, provides $0.2\msun\yr^{-1}$
of gas.

The parameter $\alpha$ has the job of quantifying the rate at which the mass
of cold gas associated with an ejected cloud increases over time, and by
conservation of momentum it predicts the rate of decrease of the velocity
centroid of the cold gas. The detailed hydrodynamical simulations of
Marinacci et al.\ (2010,2011) predict these same quantities from ab-initio
physics. Remarkably the value of $\alpha$ required to fit the LAB data
provides excellent fits to plots of mass and velocity versus time from the
simulations. This result makes sense only if the model has correctly
captured the basic physics of the Galactic fountain.

\section{Conclusions}\label{sec:conclude}

Star-forming galaxies such as our own must accrete intergalactic gas. Our
current understanding of why cool-core galaxy groups and clusters are
deficient in star formation makes it paradoxical that the only galaxies able
to accrete are those hosted by less deep gravitational potential wells in
which the coronal gas radiates too weakly to be detected. It is also strange
that in these galaxies accretion does not occur at the centre, where the
pressure must be highest and the cooling time shortest, but kiloparsecs away
in a disc. A promising resolution of this paradox is that galactic fountains
reach up and grab coronal gas, bringing it down into the star-forming disc. The
key to extracting gas from the corona far from the centre is the provision of
``seed'' cool gas, which lowers the cooling time of the coronal gas with
which it mixes in the wake of each fountain cloud.

If this model is correct, galaxies which lose their cool-gas discs, for
example in a major merger, will not be able to reform them from coronal gas;
their only hope of continued star formation is accreting a substantial body of
cool gas, for example from a gas-rich dwarf. A galaxy
that does not have
a cool-gas disc will not drive a fountain, and without a fountain gas can be
cooled out of the corona only at the centre, where it will feed the central
black hole and stimulate it into an outburst that will reheat the corona and
prevent further cooling.

Observations of the \hi\ halos of external galaxies seem to require
interactions between fountain clouds and coronal gas similar to those
predicted by basic physics: clouds must be exchanging mass and momentum with
the corona. The data require that the clouds gain mass, while hydrodynamic
simulations clearly predict that clouds are steadily ablated by the corona.
However, for an astronomically plausible range of parameters, condensation of
coronal gas in the turbulent wake of each cloud yields the predicted changes
in the total mass and mean velocity of the cool gas. The rate of accretion
predicted in this way is in excellent agreement with that required to sustain
star formation, and there is remarkable agreement between the rates of mass
and momentum exchange required by the observations and hydrodynamical
simulations.

In summary, it seems likely that supernova-driven fountains provide the
dominant mechanism through which disc galaxies collect gas from intergalactic
space, at least at redshifts $z\lsim1$.

\end{document}